\documentclass[useAMS,twocolumn,usenatbib]{mn2e}

\usepackage{graphicx}

\def\cf{{cf.~}}
\def\ie{{i.e.,~}}
\def\eg{{e.g.,~}}

\newcommand{\etal}{et~al.\ }

\newcommand\beq{\begin{equation}}
\newcommand\eeq{\end{equation}}
\def\gtsima{$\; \buildrel > \over \sim \;$}
\def\ltsima{$\; \buildrel < \over \sim \;$}
\def\prosima{$\; \buildrel \propto \over \sim \;$}
\def\gsim{\lower.5ex\hbox{\gtsima}}
\def\lsim{\lower.5ex\hbox{\ltsima}}
\def\simgt{\lower.5ex\hbox{\gtsima}}
\def\simlt{\lower.5ex\hbox{\ltsima}}
\def\simpr{\lower.5ex\hbox{\prosima}}
\def\la{\lsim}
\def\ga{\gsim}
\def\ie{{\frenchspacing i.e. }}
\def\eg{{\frenchspacing e.g. }}

\def\HI{\hbox{H~$\scriptstyle\rm I\ $}}

\def\HeI{\hbox{He~$\scriptstyle\rm I\ $}}
\def\HeII{\hbox{He~$\scriptstyle\rm II\ $}}

\def\CIV{\hbox{C~$\scriptstyle\rm IV\ $}}
\def\SIV{\hbox{Si~$\scriptstyle\rm IV\ $}}
\def\OVI{\hbox{O~$\scriptstyle\rm VI\ $}}
\def\nHI{{\rm HI}}

\def\nHII{{\rm HII}}

\def\nHeII{{\rm HeII}}
\def\nHeIII{{\rm HeIII}}

\def\remove#1{}

\begin{document}
\title{Ultraviolet Background Radiation from Cosmic Structure Formation} 
\author[Miniati et al.]
{Francesco~Miniati$^1$\thanks{fm@MPA-Garching.MPG.DE}, 
Andrea Ferrara$^{2}$,  Simon D. M. White$^1$ \& Simone Bianchi$^3$ \\
\\
$^1$  Max-Planck-Institut f\"ur Astrophysik,
Karl-Schwarzschild-Str. 1, 85740, Garching, Germany \\
$^2$ SISSA/International School for Advanced Studies, via Beirut 2-4,
34013 Trieste, Italy\\
$^3$  Istituto di Radioastronomia/CNR Sez. di Firenze, Largo E. Fermi 5, 50125 Firenze, 
Italy
}
\date{\today}
\pubyear{2001} \volume{000} \pagerange{1} \onecolumn

\maketitle \label{firstpage}

\begin{abstract} 

We calculate the contribution to the ultraviolet background (UVB) from
thermal emission from gas shock heated by cosmic structure
formation. Our main calculation is based on an updated version of
Press-Schechter theory. It is consistent with a more empirical
estimate based on the observed properties of galaxies and the observed
cosmic star formation history.  Thermal UVB emission is characterized
by a hard spectrum extending well beyond 4~Ry.  The bulk of the
radiation is produced by objects in the mass range $10^{11-13}
M_\odot$, \ie large galaxies and small groups. We compute a composite
UVB spectrum due to QSO, stellar and thermal components.  The ratio of
the UVB intensities at the H and He Lyman limits increases from 60 at
$z=2$ to more than 300 at $z=6$.  A comparison of the resulting
photoionization rates to the observed Gunn-Peterson effect at high
redshifts constrains \ the escape fraction of ionizing photons from
galaxies to be less than a few percent.  Near 1 Ry, thermal and
stellar emission are comparable amounting to about 10 \%, 20 \% and 35
\% of the total flux at redshifts of 3, 4.5 and higher,
respectively. However, near the ionization threshold for
\HeII, the thermal contribution is much stronger. It is 
comparable to the QSO intensity already at redshift $\sim 3$ and
dominates at redshifts above 4. Thermal photons alone are
enough to produce and sustain \HeII reionization already at $z\approx
6$.  We discuss the possible implications of our results for the
thermal history of the intergalactic medium, in particular for
\HeII reionization.
\end{abstract}


\section{Introduction}

Neutral hydrogen in the intergalactic medium (IGM) produces a forest
of resonant Ly$\alpha$ absorption lines in the spectra of
high-redshift quasars. The connection of these features 
to the structure formation process 
has now been now firmly established using
N-body/hydrodynamic numerical simulations
\citep{CenApJL1994,ZhangApJ1995,MiraldaEscudeApJ1996}.  
There is consensus that the observed IGM temperature
results from a balance between photo-ionization heating and
adiabatic cooling due to the Hubble expansion. 
Such photo-heating is provided by the extragalactic
ultraviolet background (UVB) whose nature, origin and 
evolution have, therefore, been subject to considerable 
investigation.

At low redshifts, the ionization balance is consistent with a pure
power-law ionizing spectrum. Traditionally QSOs have considered as the
main sources of ionizing photon.  However, a number of recent
theoretical and observational developments are at odds with this.
\citet{KimA&A2001} find that the break in the
redshift evolution of absorbers occurs at lower redshifts than
predicted by numerical simulations using a standard QSO
ionizing background \citep{hama96}, hinting at an incomplete
description of the UVB.  This led \citet[][B01
hereafter]{BianchiA&A2001} to recompute the UVB as a 
superposition of contributions from QSOs and galaxies. 
Those authors adopted an escape fraction of ionizing photons from
galaxies,
$f_{esc}\simeq 10\%$. In spite of many theoretical
\citep{DoveApJ2000,WoodApJ2000,HaehneltApJL2001,CiardiMNRAS2002}
and observational studies 
\citep{LeithererApJL1995,HeckmanApJ2001,SteidelApJ2001}
the determination of this parameter remains highly uncertain.
In \S \ref{photrat.se} below we use recent measurements of the
Gunn-Peterson effect in high redshift quasars
\citep{BeckerAJ2001,McDonald2001ApJ,FanAJ2002} to argue for values of $f_{esc}$ as
small as a few \% for high redshift galaxies (\S \ref{photrat.se}), in
agreement with independent estimates by \citet{soto03}. These authors
set a 3$\sigma$ (statistical) upper limit $f_{esc} \simlt 4$\% using
photometry of 27 spectroscopically identified galaxies with $1.9 < z <
3.5$ in the Hubble Deep Field.

Similarly, high resolution simulations using a
standard cold dark matter model and a standard QSOs
ionizing background \citep{hama96}, produce a Ly-$\alpha$
forest lines with a minimum width significantly below that observed
\citep{TheunsMNRAS1998a,BryanApJ1999}.
This appears to require additional heat sources, for example
photo-electric dust heating \citep{NathMNRAS1999},
radiative transfer effects \citep{AbelApJ1999},
or Compton heating by X-ray background photons 
\citep{MadauApJL1999}. 

The situation around redshift $z\sim 3$ is more complicated.
Apparently the spectrum 
must be very soft, with a large break at the ${\rm He}^+$ edge. 
\cite{Songaila1998} reports an abrupt 
change of the $\CIV/\SIV$ ratio at $z\approx 3$. 
This may indicate a rapid and
significant change in the shape of the ionizing spectrum, perhaps
due to \HeII reionization. 
This is corroborated by the detection of patchy \HeII Ly$-\alpha$
absorption at similar redshifts.  In addition,
\citet{Heap00} analyzed the Gunn-Peterson He II absorption
trough at $z=3.05$ found the UVB to be characterized by a high
softness parameter\footnote{Note that $S$ is defined as the ratio of the photoionization 
rate in \HI over the one in \HeII (\S \ref{softpar.se}).}
$S\approx 800$, in contrast to the harder spectrum deduced at $z = 2.87$.
However, a more recent VLT/UVES
study \citep{KimA&A2002} using 7 QSOs finds no strong discontinuity 
for the quantity $\CIV/\SIV$
around $z=3$ and suggests that it might not be a good
indicator of \HeII reionization. These results are in agreement with 
the analysis of \citet{bosara03} who studied the properties of
metal absorption systems between redshift 1.6 and 4.4 in the spectra
of 9 Keck/HIRES QSOs.

\citet{SchayeMNRAS2000} concluded that
the IGM temperature evolution differs considerably from simple
expectations based on QSOs as sole ionizing power input. 
\citet{RicottiApJ2000} found a similar result with an analogous but
independent approach. Both analyses suggest at $z\sim 3$
the IGM temperature (of the gas at mean density) undergoes a
``sudden'' jump which is interpreted as
associated with \HeII reionization.
It is worth pointing out that \citet{RicottiApJ2000} 
recover higher temperature values and a quite smoother temperature jump than  
\cite{SchayeMNRAS2000}, whereas \citet{McDonaldetal2001} 
using similar data to Shaye et al. and a more conservative analysis did not
find the same temperature step at redshift 3 (for IGM gas overdense 
by a factor 1.4 with respect to the mean). This 
provides a measure of the possible uncertainties in these experiments.

In this paper we explore another source of UVB ionizing photons, namely
thermal emission from shock-heated gas in collapsed cosmic structures.
We find that thermal radiation, produced mainly in halos 
with temperatures between 10$^6$ K and a few $\times 10^7$ K,
is characterized by a hard spectrum with many photons above the
\HI and \HeII ionization thresholds.
As noted above we limit the
escape fraction of UV ionizing photons from high redshift
galaxies, $f_{esc}$, to a few \%.
Thermal emission provides a significant fraction of \HI ionizing 
photons at redshift $\geq 3$. 
In addition, we find that thermal emission plays a major role in the
reionization of \HeII, being comparable to the QSOs at
redshift $\sim 3$ and dominating the flux at
$z> 4$. In fact it turns out that thermal photons alone are
enough to cause \HeII reionization at $z\approx
6$. Our study is based on well understood emission
mechanisms, bremsstrahlung and line emission from
optically thin thermal plasma, so we expect our results to be robust.

Thermal emission is not the only possible source of ionizing 
photons in collapsing structures. An alternative is
inverse Compton emission by relativistic electrons
accelerated in large scale structure shocks \citep{mjkr01}.  
However, based on methods
similar to those presented in \cite{min02} we find such a component
to be negligible compared to that from QSOs, a result in agreement with 
\citet{RandallApJ2001}.

The paper is organized as follows. In \S \ref{model.se}
we present the details of our model, that is we describe how we
solve the radiative transfer equation and 
estimate the radiation 
due to QSOs, stars and, particularly, thermal emission. Our 
results are presented in \S \ref{res.se} where the relevant
features of the transmitted radiation flux are described.
Finally, in \S \ref{disc.se} we discuss 
the thermal evolution of the IGM and in \S \ref{suco.se}, 
we summarize our main results.

\section{Model} \label{model.se}
\subsection{Radiative Transfer} \label{radtra.se}

\begin{figure}
\includegraphics[width =1.0 \textwidth]{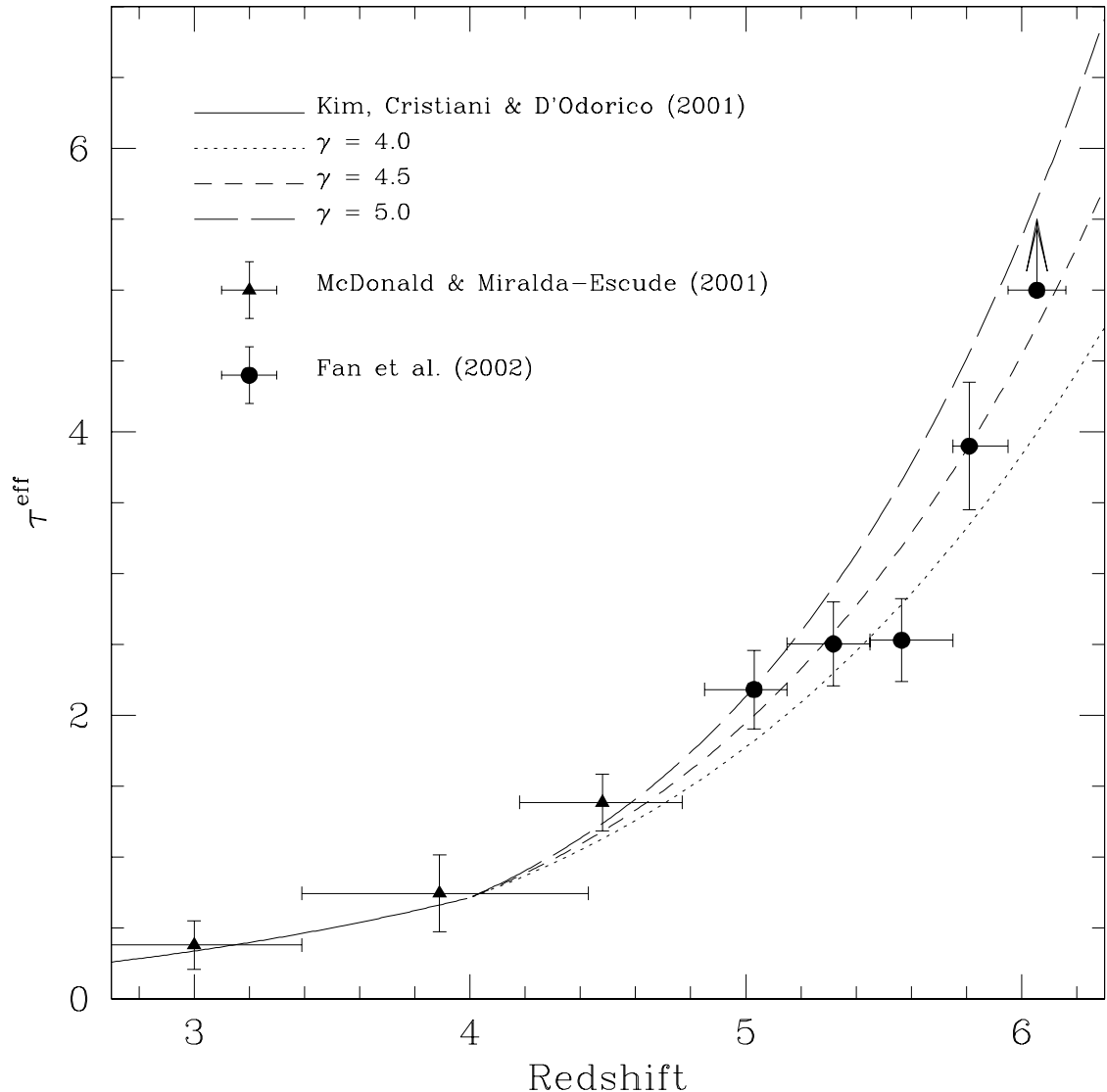}
\caption{{Multi power-law fits of the 
redshift evolution of $\tau^\mathrm{eff} 
\propto (1+z)^\gamma$. Up to $z\leq 4$ 
a single curve corresponding to $\gamma \simeq 3.4$
fits the data. For higher redshifts we show the cases 
of $\gamma=$ 5 (dot), 5.5 (dash) and 6 (long dash) 
respectively. Data point are taken from
\citet{McDonald2001ApJ} and \citet{FanAJ2002}.
} 
\label{tau.fig}}
\end{figure}

The mean specific ionizing flux, $J(\nu_o,z_o)$, observed at 
frequency, $\nu_o$, and redshift, z$_o$, 
is the solution to the cosmological radiative transfer 
equation which reads \citep{peebles93}
\begin{equation} \label{jdef.eq}
J(\nu_o,z_o)
= \frac{c}{4\pi H_0} \; \int_{z_o}^{\infty}
e^{-\tau_{eff}(\nu_o,z_o,z)} \;
\frac{[\Omega_m (1+z)^3 + \Omega_\Lambda]}{1+z}^{-1/2} \;
\left(\frac{1+z_o}{1+z}\right)^3 \;
j(\nu,z) \; dz
\end{equation}
where $c$ is the speed of light, $H_0$ is the Hubble parameter,
$\Omega_m$ and $\Omega_\Lambda$ are the matter and vacuum energy
densities respectively\footnote{ We assume a flat cosmological model
with normalized Hubble constant $h_{70} = H_0/$70 km s$^{-1}$
Mpc$^{-1}=$ 1, $\Omega_m=0.3$, $\Omega_\Lambda =0.7$,
$\sigma_8=0.9$.}, and $j(\nu ,z)$ is the volume-averaged proper
spectral emissivity computed at emission redshift, $z$, and at the
appropriately blueshifted photon frequency $\nu = \nu_o \,
(1+z)/(1+z_o)$. In addition, $\tau_{eff}(\nu_o,z_o,z)$ is the
effective optical depth at frequency $\nu_o$ due to absorption of
residual neutral gas in the IGM between $z_o$ and $z$. Following
\citet{ParesceApJ1980}, for a distribution of discrete absorbers in an
otherwise transparent (i.e. ionized) medium we write
\begin{equation}
\tau_\mathrm{eff}(\nu_o,z_o,z)= 
\int_{z_o}^z \!\!\!\!\! dz'\int_0^{\infty}
\!\!\!\! dN_{\nHI} \; f(N_{\HI},z')
(1-e^{-\tau(\nu')}).
\label{taueff}
\end{equation}
The distribution of absorbers as a function of the \HI column density
and redshift, $f(N_{\nHI},z')=\partial^2 N / \partial
N_{\nHI}\partial z'$, is typically derived from counts of Ly$\alpha$ 
lines in QSOs absorption spectra. It is assumed that 
the number density of absorbers evolves with redshift as 
$\partial N /\partial z'\propto (1+z')^{\gamma-1} $,
implying an evolution of the effective optical depth of 
the Ly$\alpha$ forest as $\tau^\mathrm{eff} \sim (1+z)^\gamma$ 
\citep{ZuoA&A1993}.
This was the approach taken in B01 who,
based on observations of the Ly$\alpha$ forest in 
the redshift range $1.5 < z < 4$ \citep{KimA&A2001},
determined a value of $\gamma \simeq 3.4$. 
For redshifts higher than 4, however, recent observations, including
the detection of a nearly
complete absorption through in the QSOs spectra
discovered by the SDSS at $z\sim 6$,
imply a much stronger evolution of that function 
\citep{BeckerAJ2001,FanAJ2002}. 
Therefore, for $z> 4$ we retain the power-law behavior 
for $\tau^\mathrm{eff}$ but determine the 
power-law index $\gamma$ through a comparison
of the power-law models with the new measurements of \citet{FanAJ2002}.
In Fig. \ref{tau.fig} we report the power-law models for different 
values of $\gamma$ together with observed data points taken from 
\citet{McDonald2001ApJ} and \citet{FanAJ2002}. Actually
the data do not exhibit a smooth behavior and, therefore, cannot be 
described in full detail by any of the smooth power-laws. The 
features in the data are likely due to inhomogeneities in the 
reionization process or to a drop in the ionizing flux in the 
aftermath of cosmological reionization \citep[\eg][]{CenMcDonald2002}. 
In addition, as we approach 
$z\sim 6$, the data indicate a sudden increase of the IGM optical depth.
Both of these features in the redshift dependence of $\tau^\mathrm{eff}$
are reflected in the redshift evolution of the ionization rates that
\citet{McDonald2001ApJ} and \citet{FanAJ2002}
inferred by modeling the distribution of absorbers 
in the IGM in a fashion similar to ours. 
Here we adopt a single power-law with $\gamma=5.5$ at $z>4$,
which approximately describes the behavior of $\tau^\mathrm{eff}(z)$.
However, it is part of our objectives to compare the 
measured ionization rates \citep{McDonald2001ApJ,FanAJ2002} with those produced
by the radiation processes explored in this paper. We will heretofore
need to bear these approximations in mind in order to 
properly interpret the results.

Each absorber is characterized by an optical depth
$\tau(\nu')=N_{\nHI}\sigma_{\nHI}(\nu')+N_{\nHeII}
\sigma_{\nHeII}(\nu')$, due to
\HI and \HeII ionization (\HeI ionization is
negligible). 
The value of $N_{\nHeII}$ can be
derived from $N_{\nHI}$ by studying the radiative transfer within each
cloud \citep{hama96}. For simplicity, we have assumed that all
clouds are optically thin at the \HeII ionization threshold, which yields
$N_{\nHeII}/N_{\nHI} \,\approx\, 1.8 \, J(13.6 {\rm eV})/J(54.4{\rm eV})$
\citep{MadauApJL1991}. While this approximation slightly
underestimates the optical depth at $h\nu > 54.4$ eV, 
it does not severely affect the opacity of the most abundant 
($N_{\nHI}<10^{17}\,\mbox{cm}^{-2}$) clouds for $h\nu<100$ eV
\citep[\cf Fig.~2 of][]{hama96}.  

In order to compute the mean specific ionizing flux 
eq.~(\ref{jdef.eq}) is solved iteratively, as required by the dependence of
$N_{\nHeII}/N_{\nHI}$ on $j(\nu,z)$. 
For the 
total emissivity we take $j=j_{\rm Q}+j_{\rm G}+j_{\rm T}$, that is the 
sum of QSO, stellar and thermal components. 
In the following sections we describe how each 
component is computed.

\subsection{Radiation from QSOs and stars in galaxies} \label{qsoga.se}

The contributions of QSOs and stars to the ionizing UV 
background adopted in this paper are similar to those derived in B01.
We briefly summarize here the main assumptions of the model.
The QSO emissivity assumes a luminosity function
that follows the double power-law model of \citet{BoyleMNRAS1988}. 
For $z\leq3$, we adopt the parameters given in 
\citet{BoyleMNRAS2000}, obtained by fitting a sample of over 
6000 QSOs with $0.35 <z < 2.3$. At $z>3$, we include the exponential 
decline suggested by \citet{FanAJ2001},
describing the dramatic reduction in QSOs number density at high
redshift. Finally the QSO spectrum in the ionizing UV range 
is modeled as 
a simple power law, $j(\nu)\propto \nu^{-1.8}$ \citep{ZhengApJ1997}.

For the stellar component, we assume a star formation rate that 
is constant from high redshifts to $z\approx1$, then rapidly 
decreases to local values, as indicated 
by several galaxy surveys in the rest-frame non-ionizing UV
\citep{MadauApJ1998,SteidelApJ1999}. Synthetic galactic spectra 
(produced with the 2001 version of the \citealt{BruzualApJ1993} code)
have then
been used to calculate the emissivity for the ionizing UV as a function 
of $z$. Finally, the internal absorption of radiation by the galaxy 
interstellar 
medium was modeled by a redshift-independent value for 
$f_\mathrm{esc}$, the fraction of Ly-continuum photons that 
can escape into the IGM. As we shall show in \S \ref{res.se}
the values for the ionization rates recently measured by
\citet{McDonald2001ApJ} and \citet{FanAJ2002}
constrain $f_\mathrm{esc}$ to a level of a few \%.
In the following we will usually set $f_\mathrm{esc}=1\%$. 
We should also mention for completeness that
we have neglected contributions from 
radiative recombinations in the absorbers \citep{hama96}.
It would be possible to correct for this omission, but it is 
not strictly necessary for comparing the contributions from 
the various emission processes explored in this paper. 
\subsection{Thermal Emission} \label{them.se}
In this section we compute the contribution to the ionizing 
background flux from thermal emission from gas
accreting onto dark matter halos. 
Since we calculate this for the first time here, we outline 
the assumptions of the model in more detail than for
the QSOs and stellar components (\S \ref{qsoga.se}). 

The gravitational collapse of matter density 
enhancements during structure formation
drives supersonic flows which eventually
shock \citep[\eg][]{minetal00}.
The shocked gas is collisionally ionized and 
heated to temperatures $T \simeq 10^6~{\rm K}~ (v/100 
~{\rm km~§ s}^{-1})^2$ so that dense regions can efficiently 
radiate away thermal energy through bremsstrahlung and line 
emission \citep{whre78}. At these temperatures and 
where the gas overdensity is above 
a few hundred 
collisions ensure the fractions,
$n_{\nHI} / n_{\nHII} \sim 10^{-7}$ and
$n_{\nHeII}/ n_{\nHeIII} \sim 10^{-6}$, for neutral hydrogen 
and singly ionized helium respectively.
It can be shown that under these circumstances the column density
associated with the gas within the virial radius produces negligible 
absorption both at 1 and 4 Ry (see Appendix \ref{taus.a}).
However, for cooler structures collisional ionization is less efficient
and the gas becomes optically thick for virialization 
temperatures $T\sim 10^5$ K (see Appendix \ref{taus.a}).  
Most of the thermal emission, however, is produced in objects 
with $T\geq 10^6$ K. . Therefore, in the following
we will assume that thermal radiation 
is characterized by an escape fraction of order 1.

In the following, the description of the evolution of the baryonic gas,
responsible for the emission of thermal radiation, is rather
simplified. 
Although in principle it could be pursued in some detail 
through semi-analytic schemes 
\citep{whfr91,kawhgu93,vdb02}, the latter provide considerably
more information than we require and their level of sophistication is
well beyond the scope of our current study.  However, we do test
our model predictions for the amount of cooling
that must have occurred against
independent empirical estimates. These are based on the observed cosmic star
formation history and the distribution of stellar mass
as a function of halo masses, that we infer from recent SDSS 
data. The comparison is detailed in \S
\ref{sdss.se}.  In addition, 
we comment on possible additional effects neglected in our
simplified approach. In general, however, the agreement that
we find between our predictions and our empirical estimates is
encouraging and supports our adopted approach.
\subsubsection{Collapsed Halos} \label{colhal.se}

The volume averaged comoving thermal emissivity in units of 
erg s$^{-1}$ cm$^{-3}$ Hz$^{-1}$ can be written as
\begin{equation} \label{avem.eq}
j_{\rm T}(\nu,z) = \int dM\; 
\frac{dn}{dM} (M,z) \;
M\,
\epsilon[T(M),\nu] \; 
\min\left[1,\frac{\tau_{cool}(M,z)}{\tau_{Hubble}(z)}\right],
\end{equation}
where $dn/dM$ is the comoving number density 
of collapsed dark matter halos of mass $M$ at redshift, $z$,
and $\epsilon(T,\nu)$ is the spectral emissivity per unit mass.
The last term is the ratio of the cooling time to the 
Hubble time at redshift $z$ and accounts for
the fact that any parcel of gas emits only as long as is
allowed by its supply of thermal energy. 

To describe $dn/dM$ we adopt the \citet{prsc74}
formalism as updated by \citet{shto99}. This gives the mass function
\begin{equation} \label{psdef.eq}
\frac{dn}{dM} (M,z) = 
A \;
\left(\frac{2}{\pi}\right)^{1/2}\; \frac{\rho_0}{M^2}\;  
\left(1+\hat{\nu}^{-2p} \right) \; 
\hat{\nu} \;
\left|\frac{d\ln\sigma(M)}{d\ln M}\right|\;   
\exp\left(-\frac{\hat{\nu}^2}{2}\right)  
\end{equation}
where $\rho_0$ is the current ($z=0$) average mass density 
of the universe, $\hat{\nu}=a^{1/2}\delta _c(z)/\sigma(M)$, 
$\sigma(M)$ is the mass variance of linear 
density perturbation corresponding to mass $M$ at the current epoch
and $\delta_c(z)$ is the critical linear overdensity evaluated at 
present for a spherical perturbation that collapses at redshift
$z$. Both $\sigma(M)$ and $\delta_c(z)$ are taken from 
\citet{kisu96} and the parameters 
$(A,a,p)=(0.322,0.707,0.3)$ are as in  \citet{shto99}.
According to the spherical collapse model, for $z\gg 1$
the virialized dark matter halos are characterized by an 
overdensity $\Delta_c\simeq 18\pi^2$
with respect to the critical value, $\rho_{cr}$.
The baryonic gas is assumed to settle in approximate 
hydrostatic equilibrium within each halo at its virial 
temperature \citep[\eg][]{brno98}
\begin{equation} \label{virtmp.eq}
T(M) = 1.4\times 10^7 {\rm K}
~h_{70}^{2/3}
\left(\frac{M}{10^{13}M_\odot}\right)^{2/3}
\left(\frac{\Omega_m}{0.3}\right)^{1/3}
\left(\frac{\Delta_c}{18\pi^2}\right)^{1/3} 
\left(\frac{1+z}{5}\right) 
\end{equation}
and with a density $\rho_b=f_b\, \Delta_c \rho_{cr}$,
where $f_b$ is the halo baryonic fraction.

The spectral thermal emissivity, $\varepsilon(T,\nu)$, 
is computed through
the code by \citet[][version 1992]{rasm77}\footnote{The code
assumes the emitting plasma in collisional equilibrium and 
neglects the effects of an ambient radiation field. 
It includes the following atomic processes: 
collisional ionization, collisional excitation followed by
auto-ionization, radiative recombination, and dielectronic recombination.
Collisional ionization and excitation are assumed to be produced by free
electrons with a Maxwellian distribution of energies \citep{rasm77}.}
whereas the cooling function, $\Lambda(T,Z)$, necessary to compute
the cooling time, was taken from the work of 
\citet{sudo93}. Our choices are appropriate because, for 
regions characterized by an overdensity of order 100 
or greater at redshift $\sim 2-3$ and higher, the
ionization equilibrium is primarily 
determined by plasma collisions.
We adopt an average metallicity within the collapsed halos 
$Z=Z_\odot /20$, thought to be typical for high redshift objects
\citep{mamaco00}. 

Contributions to the integral in eq. 
(\ref{avem.eq}) are limited below a certain mass threshold. 
For halo masses below 
$
M_{lc} = 
2.6 \times 10^{11} \; h_{70}^{-1}  \; 
(\Omega_m/0.3)^{-1/2} \; 
[\varepsilon_{13.6eV}/(1+z)]^{3/2} \;
{\rm M}_\odot,
$
where $\varepsilon_{13.6eV}$ is the photon energy in units of
13.6 eV, emission of thermal radiation is depressed exponentially 
because the low virial temperature is insufficient to ionize hydrogen.
The most stringent constraint for low mass objects is due, however, to cooling.
In fact, we find that the cooling 
correction in eq. (\ref{avem.eq}) is
\begin{equation} \label{trat.eq}
\frac{\tau_{cool}}{\tau_{Hubble}}
\equiv \alpha \frac{k_B T \tau_{Hubble}^{-1}}{n_{gas}\Lambda(T,Z)}
\simeq \alpha \, 0.4 \,
h_{70}^{-1/3} 
\left(\frac{M}{10^{13} M_\odot} \right)^{2/3} 
\left(\frac{1+z}{5} \right)^{-1/2}
\left(\frac{\Delta_c}{18\pi^2} \right)^{-2/3} 
\left[\frac{\Lambda(T,Z=Z_\odot/20)}{10^{-23} {\rm erg~ cm}^3 {\rm s}^{-1}} \right]^{-1}
\left(\frac{\Omega_m}{0.3} \right)^{5/6} 
\left(\frac{f_b}{0.15} \right)^{-1},
\end{equation}
where $n_{gas}$ is the gas number density,
$Z$ is the gas metallicity,  $\Lambda$ is the cooling function and $k_B$ is 
Boltzmann's constant.
For $Z\simeq Z_\odot/20$,
the normalization for $\Lambda$ given in eq. (\ref{trat.eq}) holds to a good 
approximation for $10^{5.5}\leq T \leq 10^{7}$ \citep[\cf][]{sudo93}.
If the gas contraction induced by radiative cooling occurs 
isobarically 
the amount of radiated energy equals the {\it enthalpy} of the system and,
$\alpha=\gamma_{gas} /(\gamma_{gas}-1)=5/2$. 
We point out that the halo life-time, 
instead of the Hubble time,
should perhaps be used in eq. (\ref{avem.eq}), but we will neglect 
these details in the current investigation.
Thus, the above scenario should 
provide at least a lower limit to the potential amount of energy 
to be radiated (even without considering the effects of feedback, see below). 

\subsubsection{Cooling \& Feedback} \label{cofe.se}

It is well known that feedback of energy and momentum associated with
formation of stars plays a fundamental role in regulating the dynamics
of the IGM, although a clear and coherent understanding of how is still 
lacking. 
For the purposes of the present investigation, the effects of
feedback can be taken into account by computing
the increase in the cooling time, $\tau_{cool}$, caused by the injection
of additional energy.
The average energy deposited in each halo of mass $M$ can 
be computed through two quantities: the
fraction $f_*$ of the baryons ($f_bM$) that is converted into stars and, 
the amount of energy, $\zeta_{SN}$, released by supernova explosions
per unit of stellar mass formed.
When normalized to the halo volume-integrated enthalpy, 
$W=(5/2)nk_BTV=(5/2)f_b Mk_BT/\mu m_p$, where $V$ is the halo volume
and $\mu m_p$ is the gas mean molecular weight, the deposited  energy is 
\begin{equation} \label{dew.eq}
\frac{\Delta E_{FB}}{W} = \frac{f_* f_b M \zeta_{SN}}{W} \simeq 2.25
\left(\frac{f_*}{0.15}\right)
\left(\frac{\zeta_{SN}}{10^{49}{\rm erg~ M}_\odot^{-1}}\right)
\left[\frac{T(M)}{10^6 K}\right]^{-1}.
\end{equation}
The value assumed for $\zeta_{SN}$ varies in the literature. If we
take a ratio of supernova per total mass converted into stars $\sim
4\times 10^{-3}$, as suggested by a standard Salpeter initial mass
function, then $\zeta_{SN} \sim 4\times 10^{48}$ erg M$^{-1}_\odot$,
although based on metallicity arguments \citet{silk97} suggests a
value 4 times larger \citep[see also][]{bookbinderetal80, ceos92}.  We
notice, incidentally, that the deposition of the assumed amount of
energy corresponds to $ \mu m_p f_* \zeta_{SN} \sim 1$ keV/part. roughly
as observed in the core of small groups of galaxies \citep{pocana99}.
Obviously, for the assumed parameters, the energy injection
will only substantially 
affect gas within halos with temperatures $\le {\rm few} \times
10^6$~K. This result is in agreement with the more sophisticated feedback
prescriptions presented in \cite{whfr91}.

With the additional energy expressed in eq. (\ref{dew.eq}) the
radiative lifetime of a halo, as in eq. (\ref{trat.eq}),
will be extended by a factor $(1+\eta \Delta E_{FB}/W)$. 
Here we have introduced yet another
parameter, $\eta(M)$, the fraction of energy deposited
through feedback processes that is available for conversion into
thermal radiation. We do so in order to account for 
various effects produced by feedback. 
For example, when $\Delta E_{FB}/W \geq 1$ gas is more
likely to be blown out of its host halo, thus strongly 
inhibiting cooling \citep{macfe99}.
This would imply $\eta \leq 0$.
In addition, the evolution of a cooling core will be altered and 
possibly disrupted by the occurrence of merger events
\cite[\cf][]{whre78}, making $\eta < 1$. 
However, as we shall show in the following, most of the radiation 
is contributed by halos within a narrow mass range. Thus, in practice, 
we only explore the case $\eta\simeq 1/2$,
which we consider appropriate for the mass range of interest here
\cite[\cf][]{mofema02}.

\subsubsection{Halo Emissivity Distribution} \label{.se}

\begin{figure}
\includegraphics[width =1.0 \textwidth]{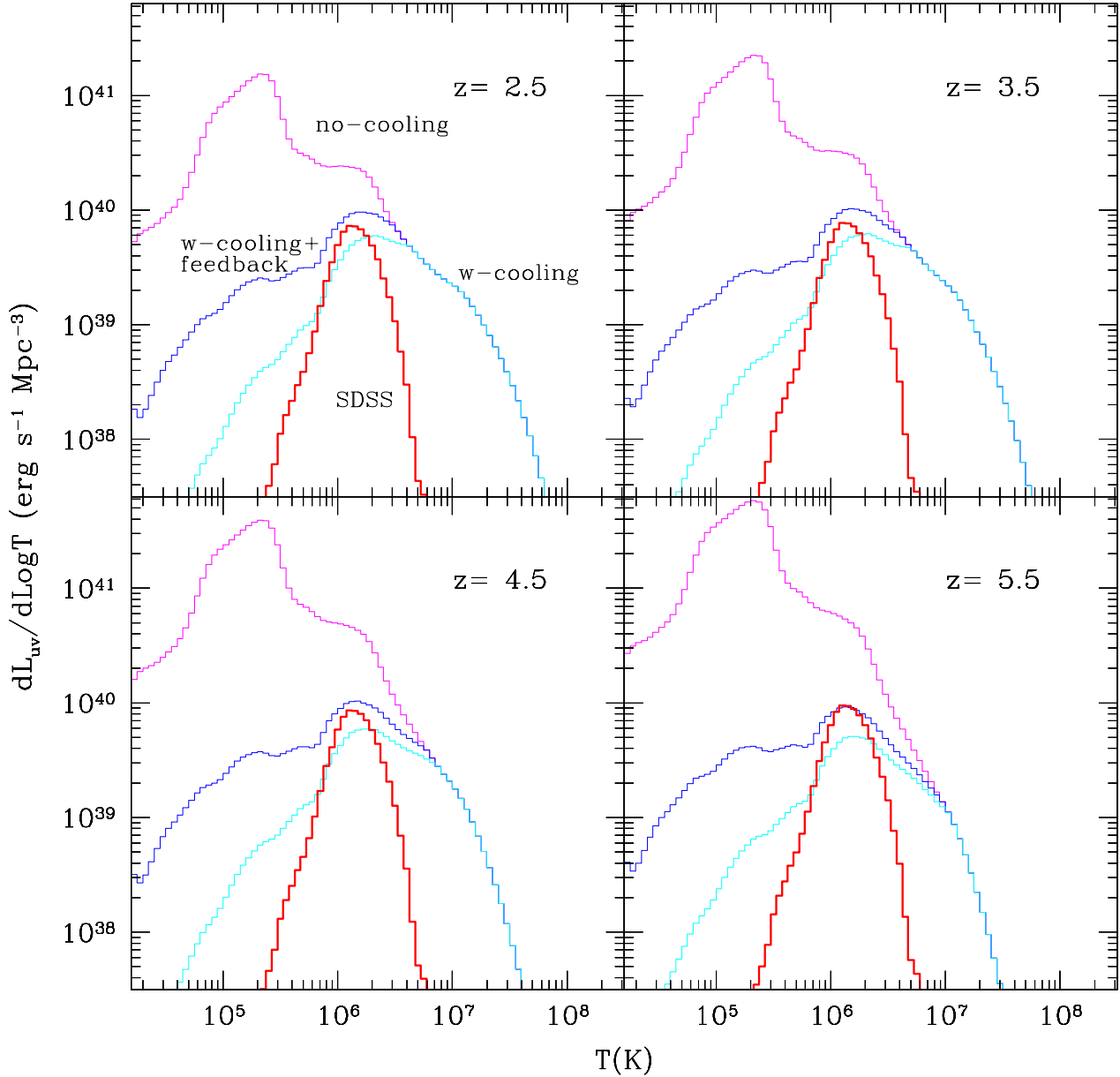}
\caption{{Histograms of the integrated thermal 
UV emission between 13.6 and 100 eV as a function of the 
virial temperature of the emitting halo population for four 
different redshifts. The various lines correspond to
cases with (w-cooling) and without (no-cooling) accounting for the 
effects of radiative 
cooling, and with cooling+feedback (w-cooling+feedback). The thick line
(SDSS) was obtained based on the observed cosmic star formation history
and the 
distribution of stellar mass as a function of halo 
virial temperature as reconstructed from SDSS data 
(see Sec. \ref{sdss.se} for details).}
\label{hiuvem.fig}}
\end{figure}

In Fig. \ref{hiuvem.fig} we show which halos are primarily responsible
for the production of thermal far UV emission. For this purpose we define
arbitrarily a UV luminosity $L_{uv}(T)$, as the integral between 13.6
and 100 eV of $j_T(\nu,z)$ given in eq. (\ref{avem.eq}).  
We then plot for different redshifts and as a function of the temperature
of the emitting halo population, histograms of $dL_{uv}/d\log T$, that is
the total UV luminosity of halos with virial temperature in a logarithmic 
interval centered on $T$.
According to our model, most of the emission 
is produced by halos with temperatures between 10$^6$ K and a
few $\times 10^7$ K, corresponding to masses $10^{11-13}$
M$_\odot$. The contribution from halos below this range is reduced
because of their short cooling time.  This may be seen by comparing with
the highest (no-cooling)
curve which corresponds to the case in which the 
corrective term that
accounts for the effects of cooling is omitted.  Notice that the
``bumpy'' shape of these curves for $T\leq 10^7$ K is a reflection of
the structured temperature dependence of the cooling function. On
the other hand, the bend at the high mass end is solely due to the
paucity of massive halos above a certain threshold which depends on
both redshift and $\sigma(M)$: for $\sigma(M)\propto M^\beta$,
$dL_{uv}/d\log T\propto dL_{uv}/d\log M \propto M^{\beta/2-1/3}$.
The line labeled `w-cooling+feedback'
corresponds to the feedback case described in 
Sec. \ref{cofe.se}. 
As expected, the extra injected energy primarily affects
small structures which are now able to produce a larger amount of 
radiation (by reradiating the injected energy). 
Compared to the no-feedback case (no-feedback) 
the spectra 
in Fig. \ref{traflu.fig} are increased by a factor of a few in 
the low energy UV range but are basically unchanged in soft X-rays. 
This is a reflection of the fact that
feedback mostly affects smaller (colder) halos.    
In any case the objects which dominate the thermal emission
are only weakly affected by the specific feedback prescription we adopt. 
\subsection{Comparison with Observational Data} \label{sdss.se}

We will now attempt to test our model against observable quantities that are
related to the occurrence of radiative cooling.
For this purpose, we have used SDSS data  presented in
\cite{kauffmannetal03} for a representative sample of
more than 10$^5$ galaxies,
to construct
a function, $g_*(M)$, describing how the total stellar mass is
distributed over halo mass in the present day universe.
The dataset includes the following quantities of interest to
us: colors in the $g,~r,~i$ bands, stellar masses and concentration
index. 

For each galaxy, the halo mass is inferred from its luminosity
through either the Tully-Fisher or Faber-Jackson relations depending
on whether the galaxy is disk or bulge dominated, respectively.
We first divide the sample into spirals and ellipticals on
the basis of the concentration index defined as the ratio of the 
Petrosian half-light radius to the Petrosian 90\% light radius,
$C=r_{90}/r_{50}$. Thus, a galaxy is classified as spiral or
elliptical/spheroidal depending on whether or not $C>3$
\citep{shimasakuetal01}. The SDSS color indexes are converted to Johnson's
$UBVR_CI_C$ system according to the relations given by \cite{smithetal02}.
Next, for spirals we use Verheijen's (2001) relation between the
$I-$band magnitude, $M_I$, and the rotational velocity $V_{flat}$,
measured in the outer parts of the galaxy disk where the rotation
curve flattens out.
For ellipticals we employ the relation between $B-$band
magnitude, $M_B$, and dispersion velocity as determined by
\cite{bezibr96}. Temperature and velocity are then related according to
$k_BT = (1/2)\mu m_p v^2$ and temperature and mass as in eq. (\ref{virtmp.eq}).
\begin{figure}
\includegraphics[width =1.0 \textwidth]{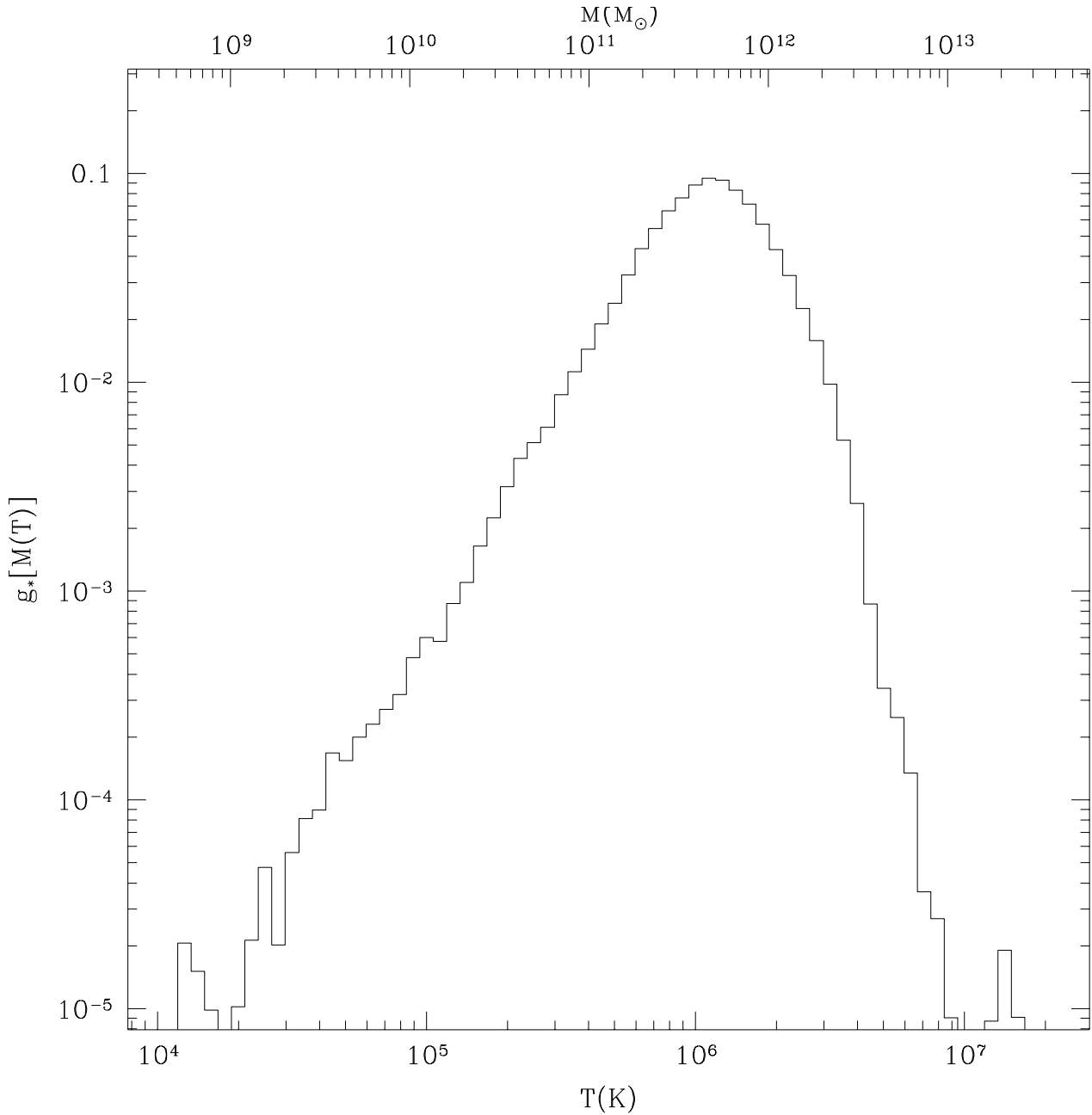}
\caption{{Histograms representing $g_*$, i.e. the fraction of 
stars in the local universe hosted in halos of virial temperature $T$,
and mass $M$. $M$ and $T$ are related through eq. (\ref{virtmp.eq})
where we set $z=0$. The histogram was built on the basis of the SDSS data 
\citep{kauffmannetal03}.}
\label{gstar.fig}}
\end{figure}
The results for $g_*$ are shown in Fig. \ref{gstar.fig}, where 
the histogram represents the fraction of stellar mass in the local 
universe as a function of the host halo virial temperature and mass.

The formation of a mass $M_*$ of stars implies that the thermal energy/enthalpy
associated with the equivalent gas mass, $\alpha M_* k T$, must have been
radiated away. Here $T(M)$ is the virial temperature of the host halo
and $\alpha=5/2$  as in eq. (\ref{trat.eq}). In fact,
this is just a lower limit. More precisely,  
the amount of radiated thermal energy is connected with the rate at which 
cold gas forms inside a halo, $\dot{M}_{cold}$, rather than the rate at which
stars form, $\dot{M}_{*}$. Obviously,
averaged over the halo lifetime $\dot{M}_{cold} \geq \dot{M}_{*}$;
however, there is no straightforward way of inferring the former from
the latter. A number of empirical facts and theoretical
arguments suggest a proportionality between the 
{\it cosmic-volume} averages of the two quantities, 
\begin{equation}  \label{mcdot0.eq}
\langle \dot{M}_{cold}\rangle =
\lambda(M,z) \langle \dot{M}_{*}\rangle (M,z)
\end{equation}
with the proportionality factor 
$\lambda(M,z)$ taking values of order of several at high redshifts
(in Appendix \ref{lambda.a} we provide an estimate of $\lambda(M,z)$ based on 
Press-Schechter formalism).
Since we only have poor statistical information about how the time
history of star formation depends on the halo mass, we assume that, on
average, stars in halos of a given mass $M$ formed at the measured
cosmic star formation rate (see, \eg B01) multiplied by our stellar
mass distribution function, $g_*(T)$. 
That is 
\begin{equation}
\langle \dot{M}_{*}\rangle (M,z) = g_*(M) \; {\rm SFR}(z).
\end{equation}
This is the simplest assumption
compatible with the results in Fig. \ref{gstar.fig}.
Thus, the differential amount of energy
radiated as a consequence of star formation by halos of mass $M$ is
\begin{equation}
\frac{d\dot{\cal E}}{dM}(z,M) = 
\alpha \frac{k_B T(M)}{\mu m_p} \; 
\langle \dot{M}_{cold}\rangle = 
\alpha \frac{k_B T(M)}{\mu m_p} \;
g_*(M)\; \lambda(M,z) \; {\rm SFR}(z).
\end{equation}
which amounts to a differential volume-averaged thermal emissivity
\begin{equation} \label{sdsslum.eq}
\frac{d j_T}{dM}(\nu,z) = \frac{d\dot{\cal E}}{dM}(z,M) \; 
\frac{\epsilon[T(M),\nu]}{\Lambda[T(M),Z]} .
\end{equation}
The last term in the above equation 
describes the spectral distribution of the radiated energy.
Using the above `empirical' thermal emissivity, and the 
cosmic star formation rate $SFR(z)$ summarized in B01,
we have recomputed the quantity $dL_{uv}/d\log T$ presented 
in Fig. \ref{hiuvem.fig} in order to compare it with our
model predictions. 
The results are illustrated by the thick (SDSS) line plotted in
Fig. \ref{hiuvem.fig} for each redshift  (the adopted value 
of $\lambda$ was computed as detailed in Appendix \ref{lambda.a}). 

The comparison between model and `empirical' predictions 
is only meaningful for halos with $\tau_{cool} < \tau_{Hubble}$
(no star formation can occur otherwise), i.e. with  
virial temperatures somewhat below the overlap point of the 
curves labeled as `with-cooling' (cooling
accounted for) and `no-cooling' (no cooling) in Fig. \ref{hiuvem.fig}.
The actual dividing line should 
occur a few times below this point of overlap,
typically above several $\times 10^6$ K.  This is because the cooling
suppression factor in the former curve is based on the assumption that
the halo age equals the Hubble time and is thus a bit overestimated.
With this clarification, we conclude that the `empirical' curve agrees
quite well with the model predictions for star forming halos with virial
temperatures above a few $\times 10^5$ K, where the cooling effects
are strong. Below $\times 10^5$ K the agreement worsens
considerably.  This is somewhat expected and is tipically attributed
to the suppression of star formation and/or gas blow-away due to feedback
effects.  Nevertheless, this is not too worrisome since the
contribution of these objects to the UV background is not crucial,
even in the `feedback-case'.  

Finally, we note that the model and empirical curves have been
derived using completely different methods and assumptions,
with values for the few free parameters involved (\eg $\theta$)
taken from the studies in which these were
introduced, rather than in order to improve our results.  Nevertheless,
the two approaches give very similar estimates of the UV thermal
emission; this is encouraging as, in principle, the agreement could have
been much worse.
\section{Results} \label{res.se}
\subsection{Ionizing Flux}
\begin{figure}
\includegraphics[width =1.0 \textwidth]{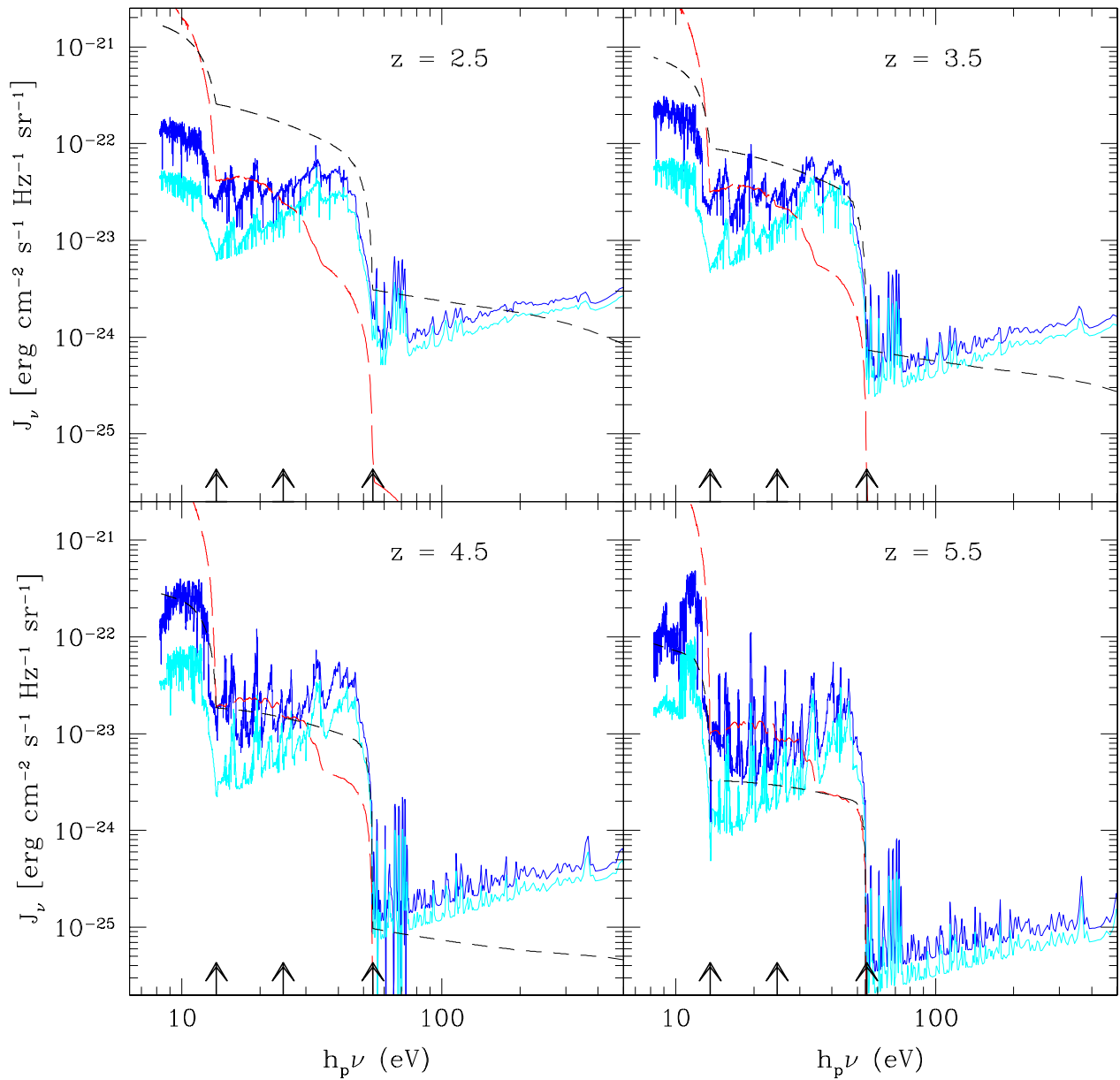}
\caption{{Mean ionizing flux due to:
thermal emission with (higher, solid) and without (lower, solid) 
feedback effects,
QSOs (dash) and stellar 
(assuming $f_{esc}= 1\%$, long-dash) for four different 
redshifts. From left to right, the vertical arrows close to 
the horizontal axis indicate ionization thresholds of \HI, \HeI and 
\HeII respectively.}
\label{traflu.fig}}
\end{figure}
\begin{figure}
\includegraphics[width =1.0 \textwidth]{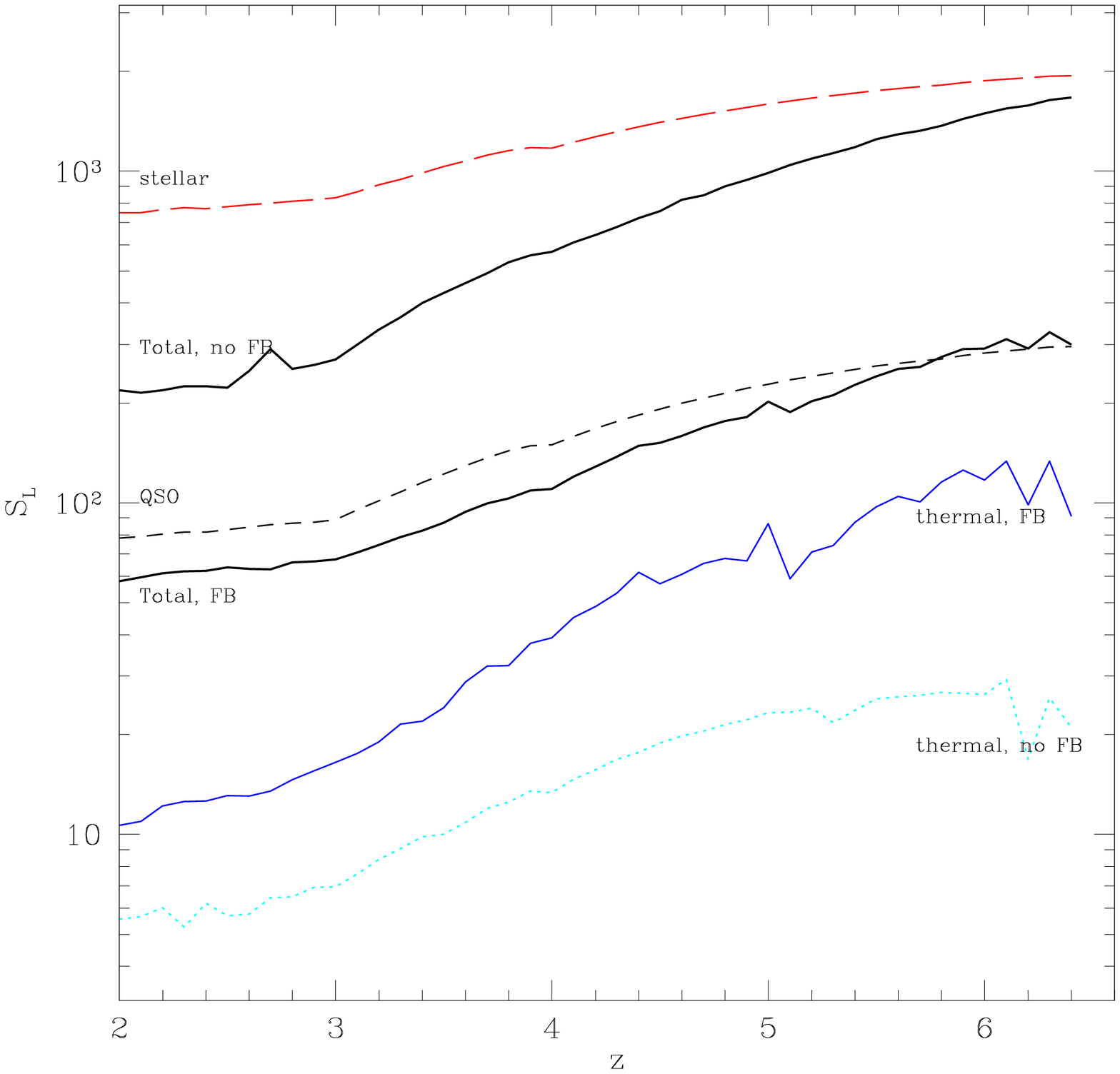}
\caption{{Redshift evolution of the softness ratio $S_L$ for 
different emission cases: thermal, no feedback (dot);
thermal, with feedback (thin solid); stellar (long dash); 
QSOs (short dash); total, no feedback (higher thick solid);
total, with feedback (lower thick solid).
} 
\label{soft.fig}}
\end{figure}
Fig. \ref{traflu.fig} shows the mean ionizing flux due to thermal
emission with ($\eta=0.5$) and without feedback effects, that due to
QSOs, and that due to galaxies ($f_{esc}= 1\%$) at four different
redshifts.  The imprint of the \HI and \HeII IGM absorption, with a
significant reduction of the flux close to the Lyman limits of the two
species, is clearly visible in each spectrum.  We find that thermal
emission provides an important contribution to the average UV
radiation flux. Apart from this, the three sources of radiation have
very different spectra and redshift dependence. At $z\sim 3$ and near
the \HI ionization threshold, thermal emission with feedback effects
corresponds to $\sim$ 10 \% of the QSO contribution and is comparable
to the stellar component. When feedback effects are not included the
thermal flux in this spectral region is reduced by a factor of a
few. At higher redshifts the thermal flux becomes progressively more
important relative to the QSO component, whereas its relation to the
stellar component is virtually unchanged.  This latter feature makes
sense because the production of stellar photons is linked to the
formation of stars and, therefore, of cold gas, which in turn produces
the thermal photons. Since the two components were modeled
independently, this provides a nice consistency check. At photon
energies near the \HeII ionization threshold the trends are similar,
except that now the thermal flux plays a more dominant role because of
its relatively hard spectrum. Thus, in this energy range, and
independently of feedback effects, thermal emission dominates the
stellar contribution (even for $f_\mathrm{esc}\sim 10\%$).  It is
comparable to the QSOs flux at $z\sim 3$, and becomes the dominant
source at redshifts $\geq 4$.
\subsection{Softness parameter}
\label{softpar.se}
Both for the individual spectral components and for the total
spectrum, we have computed the redshift evolution of the spectral
softness parameter,
\begin{equation}
S_L\equiv {J_\nHI \over J_\nHeII},
\end{equation}
where $J_\nHI$ and $J_\nHeII$ are the UVB intensities at the \HI and
\HeII Lyman limits respectively. These are shown in Fig.
\ref{soft.fig}. As already anticipated, the thermal emission is characterized
by small values of $S_L$, ranging from a few at low redshift to 100, at
$z \approx 6$, with minor differences introduced by the adopted
feedback prescription.  A similar evolution is seen for the QSO
spectra, although with higher $S_L$ values in the range
80-300; the stellar component instead
shows less pronounced evolution, increasing by only a factor of
$\sim 3$, but maximal $S_L$ values that exceed 1000 at $z \simgt
3$. The composite spectra (feedback case) $S_L$ evolution resembles
very closely that of QSOs, even at high redshift: this is somewhat
fortuitous as at $z\simgt 4$ the UVB is dominated by the sum of stellar
and thermal contribution.  The above values can be compared with the
available data. \citet{Heap00}, from an analysis of the quasar
Q0302-003 ($z=3.286$), find that around $z=3.2$
the ratio of the \HI to \HeII photoionization rates,
is larger than 400.  As
$S_L=r^{-1}\, \Gamma_{\rm H\scriptstyle I}/\Gamma_{\rm He\scriptstyle II}$, 
where $r$ depends on the spectrum
shape and is about 4 for our composite spectrum, from
Fig. \ref{soft.fig} we find that this value corresponds to our
estimate for $z=3.4$, a very close match.
In our model the results of Heap et al. would not be necessarily 
caused by a jump in the parameter $S$ at $z\simeq 3$, but by the 
evolution of the composite spectrum, particularly the decrease 
in the ratio of thermal to the QSO flux. 
In the final Section we will reconsider this point.
\subsection{Photo-ionization Rates}
\label{photrat.se}
\begin{figure}
\includegraphics[width =1.0 \textwidth]{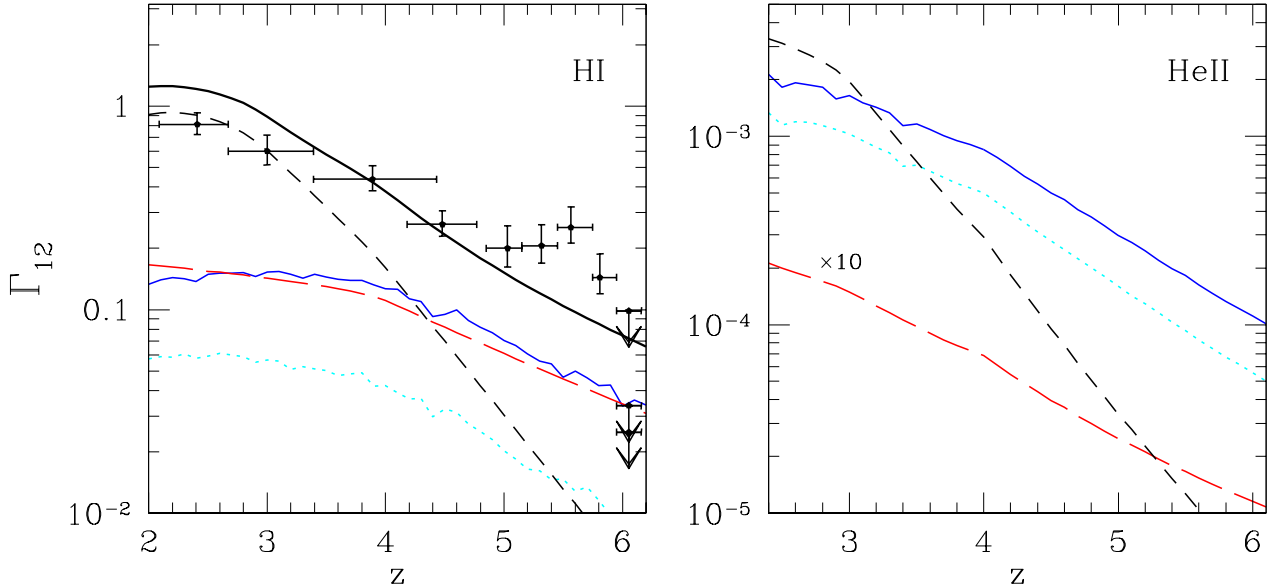}
\caption{{Left: Photo-ionization rates defined in eq. 
\ref{phion.eq} as a function of redshift for ionization of 
\HI due to emission from: QSO (dash),
stellar for $f_{esc}=1\%$ (long dash) and 
shocked IGM with feedback effects (solid) and without (dot).
The thick solid line is the total considering the feedback case
(or, alternatively, the no-feedback case and $f_{esc}=2\%$). 
The data points are from 
\citet{McDonald2001ApJ} and \citet{FanAJ2002} after correction
for our cosmological model. 
Right: same as for left panel but now for coefficients 
relative to \HeII (right). The stellar component (long dash) has
now been multiplied by a factor 10 for visualization purposes.}
\label{gammi.fig}}
\end{figure}
Fig. \ref{gammi.fig} shows the photo-ionization rates 
in units of $10^{-12}$ s$^{-1}$ defined as
\begin{equation} \label{phion.eq}
\Gamma_{12} (z) \equiv 
\frac{\Gamma (z)}{10^{-12}\, {\rm s}^{-1}} =
4\pi \;  \int_{\nu_s}^\infty \; 
\sigma_s(\nu) \; \frac{J(\nu,z)}{h_p\nu} \; d\nu
\end{equation}
for the various UV radiation components discussed above 
and as a function of redshift with the same notation as in
Fig. \ref{traflu.fig}. In addition, we also plot the 
photo-ionization rates inferred 
from the observed Gunn-Peterson effect in high-z QSO 
spectra \citep{McDonald2001ApJ,FanAJ2002}.

The left panel of Fig. \ref{gammi.fig} is relative to 
ionization of \HI and contains a number of important features. 
First, we notice that although the emission from QSOs 
is able to produce the ionizing flux observed at $z\sim 2-3$,
it falls short at higher redshifts, a well known fact. In our 
formulation, it results from the assumed rapid decline of the QSO 
number density for $z>3$, as derived from the SDSS 
\citep[\S \ref{qsoga.se}][]{FanAJ2001}. We note here, however,
that \citet{MeiksinMNRAS2003} in a recent paper questioned
this view: by tweaking
the bright end of the SDSS QSO luminosity function and allowing for
objects fainter than the survey limit, QSOs alone can in principle 
account for the UV background at $z > 4$.

As a second result, the comparison between the meadured values of
$\Gamma_{12}$
\citet{McDonald2001ApJ,FanAJ2002} and the stellar ionization rates assuming 
$f_{esc}=1\%$ (dash curve) imply, according to our model, that 
$f_{esc}$ has to be smaller than a few \%. In fact, had we assumed, \eg, 
$f_{esc}\simeq 10\%$ the observed $\Gamma$ would have been 
overpredicted by a factor 5 for $z\geq 4$ and by a factor 2 at $z\sim 2- 3$. 
In B01 the value $f_{esc}=10\%$ was preferred, based on comparisons of
the predicted ionizing flux with estimates from the proximity effect at
$2 \leq z\leq 4$. Estimates of the UV background from the proximity
effect are known to be larger than those obtained via theoretical models
of the IGM opacity (\ie the work of \citeauthor{FanAJ2002} we are using
here), probably because of a bias of the QSO distribution 
towards the denser environments the \citep{SchirberApJ2002} 
or because of systematic errors
due to line blending \citep{ScottApJS2000}. Furthermore, proximity-based
UV backgrounds have been traditionally derived in a flat Einstein-De
Sitter cosmology \citep{ScottApJS2000}: this may cause a further 
overestimation by 40\% if the {\em true} cosmology is the one we are using 
here \citep{PhillippsMNRAS2002}.

Our conclusions on $f_{esc}$ are not affected by the high values
of the measured photo-ionization rates for $5\leq z \leq 6$. 
As already pointed out in \S  
\ref{radtra.se} (see Fig. \ref{tau.fig} there) that bump 
is due to a deviation of the IGM optical depth from the
assumed smooth power-law evolution.
Had we modeled in more detail the behavior of 
the IGM optical depth we would have also recovered higher values for 
$\Gamma$ in the same redshift range. Therefore our
statement above is valid throughout the whole redshift range. 

A third interesting point concerns the relative contribution
of stellar and thermal emission to the metagalactic UV flux
at high redshifts where the QSO emission drops rapidly. 
Although one can in principle reproduce the observed
photo-ionization rates (thick solid line) 
with an escape fraction $f_{esc} \sim 2\%$,
it is also possible that at least half of those rates are due to
thermal emission. This is the case when a fraction $\eta\simeq 0.5$
of the feedback
energy is re-radiated through thermal processes (thin solid line).
However,  thermal emission is
characterized by a much harder spectrum than the stellar one. This might
serve not only as a way to discriminate observationally between
the two components, but it might have implications for the temperature
evolution of the IGM, studied
in the next Section.

Finally, the right panel of Fig. \ref{gammi.fig} shows 
the \HeII ionization rates. According to the plot, above
4 Ry stellar emission is thoroughly negligible (independent 
of $f_\mathrm{esc}$) whereas thermal emission 
is comparable to QSOs at $z\sim 3$ but completely dominates 
the radiation flux at higher redshifts. 
This result is very important in terms of the IGM evolution and 
has not previously been noticed.
It depends only weakly on feedback, but it does
assume an escape fraction of the thermal photons 
from collapsed halos of order 1. As already 
discussed in Sec. \ref{them.se} and shown in
Appendix \ref{taus.a}, these conditions should be ensured for 
collisionally ionized gas within halos of virial temperature above
$10^6$ K, that is for the halos that generate most of the thermal
emission. However, at 4 Ry thermal flux would dominate over the QSO 
component even for an escape fraction from halos $\geq 0.3$ at
$z\geq 4$ and as small as $\geq 0.05$ at $z\geq 5.5$.

\subsection{\HeII Reionization} \label{heiireion.se}
The previous results hint at the intriguing possibility that \HeII
reionization could have been powered by UV light from cosmic structure
formation. For this to be the case,
the production rate of ionizing photons has to satisfy the
condition
\begin{equation}
\Gamma \times \min(\tau_{rec},\tau_{Hubble}) \geq 1 .
\label{reccon}
\end{equation}
Here $\tau_{rec} \simeq 0.9/ \bar n_{gas} \alpha(T) C $ is the
recombination time, $\alpha$ is the radiative recombination coefficient,
and the clumping factor $C \equiv \langle n_p^2\rangle/\langle
n_p\rangle ^2 > 1$ is meant to allow for the effects of density
inhomogeneities inside the ionized region.  Using a helium
to hydrogen number ratio $y=0.08$ and assuming a temperature of the
reionized gas $T \approx 4 \times 10^4$~K,  we find
\begin{equation}
\tau_{rec} = 5.3 \times 10^{15}\; C^{-1} \left(1+z\over 10\right)^{-3} \; {\rm ~s},
\label{trecn}
\end{equation}
which is shorter than a Hubble time for $z \gsim 4.5$.  Thus, from
Fig. \ref{gammi.fig} we find that at $z\approx 6$ thermal emission
dominates the photoionization rates and alone provides $5.5\times
10^{-17}$ ($10^{-16}$) \HeII photoionizations/s in the no-feedback
(feedback) case. According to eq. (\ref{trecn}), the
\HeII recombination rate at the same redshift is $6.4\times 10^{-17}
C$~s$^{-1}$, \ie a comparable value. Hence,
it appears that structure formation can produce \HeII reionization
around $z=6$, without the contribution from any other process and
essentially independently of the feedback prescription adopted.

An additional check of the above results can be performed. At $z=6$
the fraction of mass in halos with virial temperature [calculated from
Press-Schechter, eq. (\ref{psdef.eq})] larger than 4 Ry$/k_B$ is equal
to $f_4 =3$\% \citep{MoMNRAS2002}. In order to provide at least one
\HeII ionizing photon per He atom (we have assumed a helium abundance
$y=0.08$ and that He is all singly ionized), the required mean photon
energy has to be equal to 4 Ry$\times (y/f_4) = 145$~eV. By averaging
again over the mass distribution we find that this mean energy is 1.7
keV. Hence this simple argument confirms that \HeII reionization can
be caused by structure formation.

Whether or not this possibility is fully compatible with other 
observational results is not clear. Nevertheless,
in \S \ref{disc.se} we show that this scenario predicts a temperature
evolution of the IGM that maybe consistent with existing observations.  
\section{Discussion} \label{disc.se}
\begin{figure}
\includegraphics[width =1.0 \textwidth]{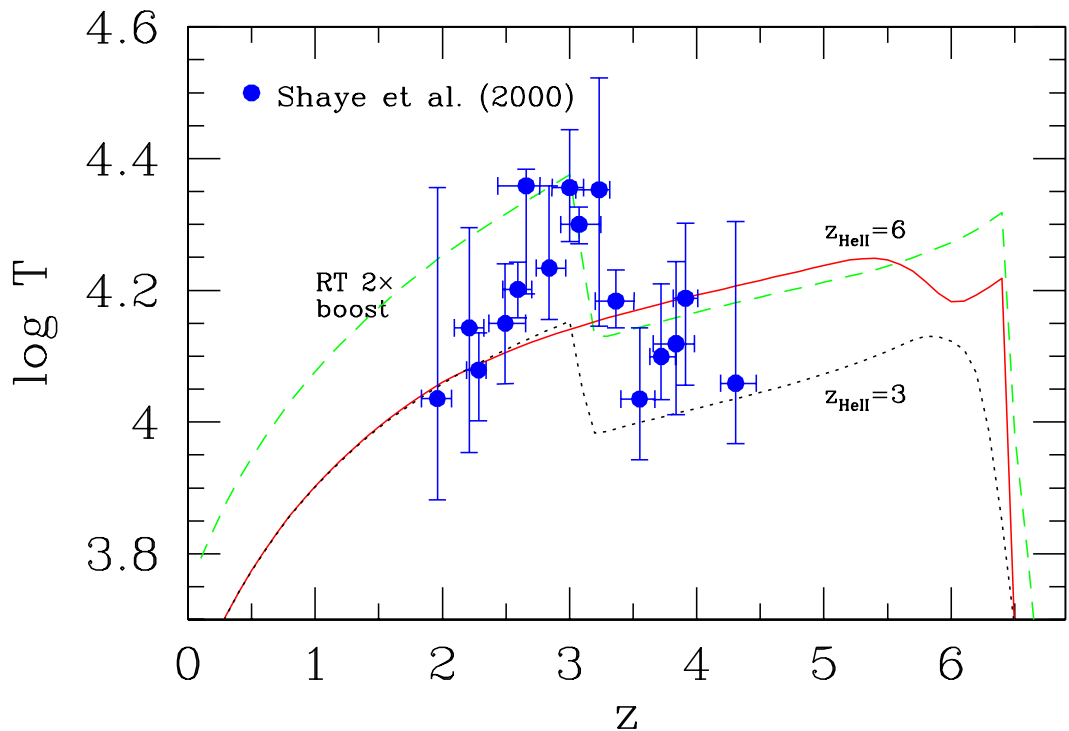}
\caption{{IGM temperature evolution:
For \HI reionization redshift $z_\nHI=6$ and
\HeII reionization redshift $z_\nHeII=3$ we show the optically 
thin case (dot) and a case in which radiative transfer effects
enhance the photoheating rates by a factor 2 (dash).
Finally the case of an early \HeII reionization at $z_\nHeII=6$
is illustrated (solid). 
An escape fraction for stellar photons $f_{esc}=1\%$ is assumed.
The data points are from Schaye \etal (2000).} 
\label{therm.fig}}
\end{figure}

Fig. \ref{therm.fig} illustrates the thermal history of the IGM.
Data points are from \citet{SchayeMNRAS2000},
who measured the cut-off in the $b - N_{\nHI}$ relation in a
sample of nine quasar spectra in the redshift range 2.5-4.0.  Despite 
the large uncertainties, there is an indication of a temperature peak at
redshift $z\approx 3$. 
On the other hand, the plotted curves represents different evolutionary
scenarios and are (numerical) solutions  to the equation
\begin{equation} \label{tevol.eq}
\frac{dT}{T} = (\gamma-1)\left[\frac{(\Gamma - \Lambda)dt}{y_t n_H k_B T}
+ \frac{dn_H}{n_H} + \frac{dy_t}{y_t} - d\frac{1}{\gamma - 1}\right] + \gamma
\frac{d\mu}{\mu}
\end{equation}
where $y_t = \sum y_i$, $y_i = n_i/n_H$ is the concentration of
the i-th species, $\mu = y_t^{-1} \sum y_i\mu_i$ is the mean molecular 
weight and the IGM mean density evolves as $n_H(z)=n_H(0)(1+z)^{3}$.
The calculation accounts for photoheating from a time dependent ionizing
background, Compton cooling on Cosmic Microwave Background photons
and non--equilibrium evolution of hydrogen and helium ionic species in a
cosmological context 
\citep[see][ for further details; we do not include metal cooling 
which is negligible for typical IGM 
metallicities]{FerraraMNRAS1996,TheunsMNRAS1998b}. 
The temperature of the medium at $z=10$, where we start our
integration, has been computed self-consistently.

We emphasize that the various curves presented in Fig. \ref{therm.fig}
are meant to illustrate a variety of possibilities and not to depict a
specific scenario. In this perspective, 
we first discuss the dotted line, representing
the case for a UVB generated by
QSO+stellar contributions given 
by the photoionization rates presented in Fig.
\ref{phion.eq}. In this case
\HI reionizes at $z\sim 6$ and \HeII around $z\sim 3$ 
(therefore their heating
and ionization rates are set to zero prior to their respective epoch of ionization).
At hydrogen reionization the IGM temperature jumps 
to $\log T = 4.13$ and decreases later on
due to adiabatic cooling. 
A sharp rise is seen at \HeII reionization at $z_{\nHeII}=3$, 
although not quite enough to fit all the data points. 
When the contribution from thermal emission is added, 
the effect is small since the
previous photoionization rates were already sufficient to keep the
gas fully ionized and, therefore, their increase 
changes little the energy input to the IGM. 

As a second example the long-dash-dot curve illustrates a similar
case but with the photoheating rates increased twofold 
to mimic radiative transfer effects as discussed in
\citet{AbelApJ1999}. Again, 
\HI reionizes at $z\sim 6$ and \HeII around $z\sim 3$.
This scenario provides a somewhat better fit. Obviously it is possible
to improve even further the agreement with the data points by fiddling
with the available parameters so as to produce the desired behavior.
At least in principle the data points can be reproduced 
\cite[see also][]{RicottiApJ2000}.

Motivated by our findings in \ref{heiireion.se} we also explore a 
less conservative scenario in which \HeII reionization occurs 
as early as $z_{\nHeII}=6$ and is mainly driven by thermal 
emission. By construction, the temperature peak around $z\sim 3$ 
does not develop in this case. Nevertheless, most of the data are 
fit at the 1 $\sigma$ level and all except one point at $2 \sigma$ level.

The possibility that cosmic structure formation is responsible for
\HeII reionization at an early epoch 
is intriguing and challenges the common wisdom that
such process is due to QSOs and occurs around $z=3$. It is worthwhile
then to revisit the arguments in support of the standard
conjecture. These are essentially three: (i) an apparent boost of the
IGM temperature around $z=3$ which could be due to heating associated
with \HeII reionization; (ii) an abrupt change of the \CIV/\SIV ratio
at $z\approx 3$ indicating a significant change in the ionizing
spectrum; (iii) the detection of a patchy \HeII Ly$\alpha$ absorption at
about the same redshift, suggestive of the final (overlapping) stages
of reionization process.  Taken together these facts seem to justify
the conjecture above.

However, as already pointed out by \citet{SchayeMNRAS2000},
these three observations probe different structures:
metal line ratios probe high density regions; the effective
\HeII Ly$\alpha$ absorption is mainly produced by neutral gas in the voids;
the IGM temperature measurements concern the ionized gas 
with density around the cosmic mean. Radiative transfer calculations
show that the ionization fronts propagate first through the low
density regions and only subsequently penetrate the 
denser clumps such as filaments and halos 
\citep{Gnedin2000ApJ, CiardiMNRAS2003}.

We also emphasize the following difficulties.  The existence of the
peak in the IGM temperature evolution needs further data to be fully confirmed. 
Although a wavelet analysis seems to confirm it
\citep{TheunsMNRAS2002a}, the data are also marginally consistent with
no feature at $z=3$, in which case early reionization would provide a
better fit to the temperature data (Fig. \ref{therm.fig}).  The
conclusion about the jump in the \CIV/\SIV ratio at $z=3$ is still
debated \citep[\eg][]{bosara98,Boksenberg1998}.  
More recent studies of metal absorption systems in 
QSOs spectra both with VLT/UVES \citep{KimA&A2002} and 
Keck/HIRES \citep{bosara03} find no 
discontinuity in the \CIV/\SIV ratio around $z=3$.
On the other hand, we have shown (Sec. 3.2) that our
smoothly increasing $S_L$ is quantitatively consistent with the
\citet{Heap00} observations.  Finally, \citet{Songaila1998} points out
that a considerable number of hard photons are required at $z>3$ to
reproduce the observed abundances of \OVI. At that epoch a sizable
fraction of cosmic volume has to be transparent at photon energies
above the \HeII ionizing threshold. 

In conclusion, there is no conclusive argument against an early
($z\approx 6$) start of \HeII reionization which may result from
the hard UV field emitted during the
virialization of large galaxies/small groups. Clearly, this
conclusion needs to be confirmed by numerical simulations, although
this task might be far from easy, as the correct treatment
of the strongly radiating cooling inhomogeneities in the gas poses
challenging numerical problems. Resorting to improved observations of
the proximity effect might then provide a check of the scenario proposed here.

\section{Summary and Conclusions} \label{suco.se}
We have shown that UVB ionizing photons
can be copiously produced by thermal emission from shock-heated gas
in collapsing cosmic structures. Our calculations are
based on an implementation of the extended Press-Schechter theory. 
However, the estimated amount of thermal radiation is consistent with
that inferred from an independent analysis based
on observed, high redshift star formation rates 
\citep[B01;][]{lanzettaetal02}, and the distribution
of stellar mass as a function of halo virial temperature
as reconstructed from recent SDSS data \citep{kauffmannetal03}.

Thermal radiation is characterized by
a hard spectrum extending up to photon energies of order $h_p \nu\sim
k_B T$. This is well above the \HI and \HeII ionization thresholds for
virial temperatures above $10^5$ K.
The bulk of the emission is produced by halos
with temperatures between 10$^6$ K and a few $\times 10^7$ K,
corresponding to masses $10^{11-13}$ M$_\odot$. 
We assume that most of the thermal radiation is able to freely
escape into intergalactic space, which is justified for a gas
that is collisionally ionized and at the temperature of these halos
(see Appendix \ref{taus.a}).

We use simplified radiative transfer to compute the transmitted flux
due to QSO, stellar and thermal emissions.  Importantly, the resulting
associated photoionization rates, when compared to measurements of the
Lyman series Gunn-Peterson effect in the spectra of high redshift QSOs
\citep{FanAJ2002,BeckerAJ2001}, imply an escape fraction of UV
ionizing photons from galaxies, $f_{esc}$, below a few \%. This result
is in agreement with very recent and independent determinations of
$f_{esc}$ carried out by \citet{soto03}, who set a 3$\sigma$
(statistical) upper limit $f_{esc} \simlt 4$\% for a sample of
spectroscopically identified galaxies of redshift $1.9 < z < 3.5$ in
the Hubble Deep Field.

It turns out that near the
\HI ionization threshold, thermal emission is comparable to the
stellar component and amounts to about 5-10 \%, 15-30 \% and
20-50 \% of the total at redshifts of 3, 4.5 and higher
respectively. The quoted range depends on the fraction of feedback
energy allowed to be re-radiated through thermal processes. 
Near the ionization threshold for
\HeII, the thermal contribution is much stronger. It is comparable to
the QSO input already at $z\sim 3$, and it dominates for
$z> 4$. Thus, this contribution, with a
typical softness parameter $S_L=10-100$, is expected
to play a major role in \HeII reionization. 
In principle
structure formation alone provides enough photons to produce
and sustain \HeII reionization at $z\sim 6$. These
conclusions are independent of our feedback prescriptions, which only affect
low virial temperature systems. In our scenario,
\HeII ionizing photons are produced primarily by relatively large 
collapsing structures (with $T_{vir} \simgt 10^6$~K). 

The thermal spectrum $J_\nu \propto \nu^{\alpha}$ is very hard, with a
slope in the range $\alpha \approx 1-2$ depending on the importance of
the supernova feedback we include. The latter process primarily
affects the smallest systems, increasing their emissivity through 
reradiation of SN input. As these smaller objects dominate
the 1-4 Ry band, the spectrum in this energy range becomes flatter as
feedback is increased. Measuring the evolution of the
thermal component of the UVB provides a powerful method to evaluate
the importance of SNe input into the intergalactic medium.

\section*{Acknowledgments}
We are indebted to X. Fan for providing us with the measurements of
the photo-ionization rates presented in his work (2002)
and to G. Kauffmann for making available 
to us her SDSS data sample. In addition, we wish to thank 
B. Ciardi, G. De Lucia, F. van den Bosch and V. D'Odorico 
for useful discussions. This work was partially supported by the Research
and Training Network `The Physics of the Intergalactic Medium',
EU contract HPRN-CT2000-00126 RG29185.

\appendix
\section{Optical Depth for Collisionally Ionized Gas in a Virialized Halo.} 
\label{taus.a}

Under collisional equilibrium the fraction of neutral hydrogen
and singly ionized helium are given by 
%
%
the ratio of the coefficients for radiative
recombination and collisional ionization.
With the values for these coefficients as 
summarized in \cite{TheunsMNRAS1998b} we find
\begin{eqnarray} \label{fnh.eq}
\frac{n_{\nHI}}{n_{\nHII}}  & \simeq & 
1.3\times 10^{-7} \; T_6^{-1.2} \; \frac{1+T_5^{1/2}}{1+T_6^{0.7}}
\; \exp\left(\frac{1.57809}{T_5}\right) \\
\frac{n_{\nHeII}}{n_{\nHeIII}} & \simeq & 
7\times 10^{-6} \; T_6^{-1.2} \; \frac{1+T_5^{1/2}}{1+T_6^{0.7}}
\; \exp\left(\frac{6.31515}{T_5}\right) 
\label{fnhe.eq}
\end{eqnarray}
where we have used the notation $T_n\equiv T/10^n$ K.
Within a virial radius 
\begin{equation}
R_v(T)= 0.14 \; {\rm Mpc} \; h_{70}^{-1} \; T_6^{1/2} \;
\left(\frac{1+z}{7}\right)^{-1/2} \;
\left(\frac{\Delta_c}{100}\right)^{-1/2} \;
\left(\frac{\Omega_m}{0.3}\right)^{-1/3} 
\end{equation}
the gas column density for absorption at 1 Ry (4 Ry)
is $N= f_n f_s n_{gas} R_v$, where 
$f_n$ is given in eq. \ref{fnh.eq} (\ref{fnhe.eq})
when considering that H (\HeII) is
nearly completely ionised, 
$f_s$ is the hydrogen (helium) abundance by number and $n_{gas} = 
(\Omega_b/\Omega_m) (\rho_{cr}\Delta_c/m_p) (1+z)^3$. 
Thus for the optical depth, $\tau = \sigma N$, we find
\begin{eqnarray} 
\tau(1~ Ry) & = & 6\times 10^{-3}\; T_6^{-0.7} \; 
\frac{1+T_5^{1/2}}{1+T_6^{0.7}} \;
\left(\frac{1+z}{7}\right)^{5/2}  \exp\left(\frac{1.57809}{T_5}\right) \\
\tau(4~ Ry) & = & 2.5\times 10^{-2}\; T_6^{-0.7} \; 
\frac{1+T_5^{1/2}}{1+T_6^{0.7}} \;
\left(\frac{1+z}{7}\right)^{5/2}  \exp\left(\frac{6.31515}{T_5}\right) .
\end{eqnarray}

\section{Press-Schechter Estimate of $\lambda(M,z)$.} \label{lambda.a}

The lifetime of a halo is 
\begin{equation} \label{tauh.eq}
\tau_{h}(z,z_f,M) \equiv \tau_{Hubble}(z) - \tau_{Hubble}[z_f(M)], 
\end{equation}
with the formation redshift, $z_f(M)$, of a halo of mass $M$ defined 
as the formation time of a progenitor with mass $M/2$ 
\citep{laco93,laco94}. 
Its relation to the running redshift $z$ (assumed $\gg 1$) is 
statistical and $M$-dependent, and is described by
\begin{equation} \label{zform.eq}
\frac{1+z_f}{1+z} = 1 + \frac{1}{1+z} \, \frac{\tilde{\omega}_f}{\delta_c(0)}
\left[ \sigma^2(M/2) - \sigma^2(M) \right]^{1/2}.
\end{equation}
where $\tilde{\omega}_f$ takes values (effectively between a few
$\times 10^{-3}$ and a few) according to a probability function 
defined within the extended Press-Schechter formalism
\cite[][see also approximation formulae in \citealt{kisu96}]{laco93} 
and $\delta_c$ and $\sigma(M)$ have been defined in \S \ref{colhal.se}.
Thus,
at a given redshift $z$ 
the formation rate of cold gas 
within a halo of mass $M$
that formed at $z_f$,
{\it averaged} over the halo lifetime
$\tau_{h}(z,z_f,M)$, is
\begin{equation} \label{mcdot1.eq}
\langle \dot{M}_{cold}\rangle _{time} (z,z_f,M) = 
\frac{M_{cold}}{\tau_{h}(z,z_f,M)} + (1+R) \langle \dot{M}_{*}\rangle  _{time}
\end{equation}
where $R$ ($\approx 1$) 
is the mass return fraction by stars and the subscript explicitly indicates 
$time$, as opposed to cosmic-volume. After introducing the
characteristic star formation timescale:
$\tau_\star \equiv M_{cold}/\dot{M}_{*}$ \citep[\eg][]{zasov95},
we rewrite eq. (\ref{mcdot1.eq}) as
\begin{equation} \label{mcdot2.eq}
\langle \dot{M}_{cold}\rangle _{time} = 
\left(1+R +\frac{\tau_\star}{\tau_{h}}\right) \; 
\langle \dot{M}_{*}\rangle _{time}
\end{equation}
It follows that at early epochs when $\tau_{h} \ll \tau_\star$
radiatively cooled gas is produced at a much faster rate than
stars. However, the two quantities converge at late times, when
$\tau_{h} \gg \tau_\star$.
This is consistent with, \eg, 
\cite{vdb02} semianalytic model of a Milky Way-like galaxy, in which 
star formation occurs at a rate proportional to the amount of available
cold gas. Based on the diagram of Fig. 2 of his paper we readily infer 
that $\lambda\simeq 5.5$, at $z \simeq 9$ shortly after the galaxy
starts evolving, $\lambda\sim 2.0$ by $z\simeq 3$ and $\lambda\sim
1.2$ by $z\simeq 0$.

Values of $\tau_\star$ $\sim$ Gyr or so have been inferred from 
observations of nearby galaxies by, e.g., \citet{zasov95},
although the author suggests 
higher values during the early stages of the evolution of galactic disks.
Alternatively, $\tau_\star$ can be taken from semianalytic models
which successfully reproduce a number of
observed galactic properties.  In the model of \cite{kawhgu93}, the star
formation rate is prescribed according to $\dot{M}_{*} = \theta
M_{cold}/\tau_{dyn}$, where $\theta= 0.1-0.3$ is a control parameter and
$\tau_{dyn}\equiv r_{gal}/v_{gal} = 0.08
\sqrt{\Omega_m/0.3} \;\tau_{Hubble}(z)$.  
In this formulation
\begin{equation} \label{taustar.eq}
\tau_\star \equiv \frac{\tau_{dyn}}{\theta} = 
0.08 \left(\frac{\Omega_m}{0.3}\right)^{1/2} \; \frac{\tau_{Hubble}(z)}{\theta}.  
\end{equation}
It implies lower values than those found by \citet{zasov95}
and, therefore, is more conservative for our present calculations.
After using this expression in eq. (\ref{mcdot2.eq})
and volume averaging over halos of a given mass $M$,
a comparison to eq. (\ref{mcdot0.eq}) leads to
\begin{equation} \label{lambdasf.eq}
\lambda(M,z)= 
 1+R+ \frac{1}{\theta}
\left<\frac{\tau_{dyn}(z_f)}{\tau_{h}(z,z_f)} \right>
\simeq
 1+R+ 0.1 \frac{\Omega_m}{\theta}
\left<{\left[\left(\frac{1+z_f}{1+z}\right)^{3/2} -1\right]^{-1}} \right>
\end{equation}
where the last equality holds for $z \gg 1$. The averaging operation
is effectively over the halo formation time which we achieve through 
the relation in eq. (\ref{zform.eq}) and the probability functions 
for $\tilde{\omega}_f$ as in \cite{kisu96}.
In fact, for $\theta \sim 0.1-0.3$, values of $\lambda$ computed 
through eq. (\ref{lambdasf.eq}) are of order of several.
This is consistent with the idea that the bulk of the cold gas in galactic
systems formed at $z > 2-3$ and, on average, had not yet been depleted 
by star formation occurring at roughly constant rate before that time     
\citep[see plot 2 in B01; see also][]{lanzettaetal02}.
For the purpose of comparison in \S \ref{sdss.se}
we have used $\theta=0.2$ and
$R=1$, although results are not too sensitive to $R$.

\bibliographystyle{apj}
\bibliography{papers,books,papigm}

\label{lastpage}
\end{document}